\documentclass[12pt]{article}
\usepackage[dvips]{graphicx}
\usepackage{epsfig,amsmath}
\begin{document}
\thispagestyle{empty}
\begin{center}

{\Large\bf{Summary on Theoretical Aspects \footnote{Invited talk at the 13th International Conference on Elastic and Diffractive Scattering, "`13th Blois Workshop"', CERN, Geneva, June 29-July 3, 2009.}}}

\vskip1.40cm
{\bf Jacques Soffer}
\vskip 0.3cm
Physics Department, Temple University\\
Barton Hall, 1900 N, 13th Street\\
Philadelphia, PA 19122-6082, USA 
\vskip 1.0cm
{\bf Abstract}\end{center}

During the five days of this conference a very dense scientific program has enlighted our research fields, with the
presentation of large number of interesting lectures. I will try to summarize the theoretical aspects of some of these new results.

\section{Introduction}
This meeting has confirmed once more a clear scientific evolution, that is the foundations of elastic scattering and diffraction phenomena at high energy, are now best understood in terms of the first principles of quantum chromodynamics (QCD). Since we are just at the start-up of the LHC, a great deal of the theoretical activity focus on the accessibility to this new energy regime, for the interpretation of relevant aspects of the strong interactions, by means of basic QCD mechanisms. Clearly they have to be confronted with experimental measurements, a very relevant part which will be summarized elsewhere~\cite{JD}.\\

I will essentially cover the following topics:

\begin{itemize}
\item Elastic and Total Cross Section
\item Soft Diffraction
\item Hard Diffraction and Central Production
\end{itemize}
Unfortunately, I have left out some important topics, in particular, ultra high energy cosmic rays and heavy-ion physics, because of lack of time and I apologize for that.

\section{Elastic Scattering and Total Cross Section}
This is a classical subject which was very largely discussed from different viewpoints. Let us first
mention a study of the amplitudes $pp$ and $\bar pp$ elastic scattering in the Coulomb-Nuclear interference (CNI) region~\cite{EF,KFK}, using
a method based on derivative dispersion relations. The real and imaginary parts of the hadronic amplitude near
the forward direction, whose detailed knowledge is needed, are parametrised by a single exponential, with two different hadronic slopes $B_R$ and $B_I$. The analysis of the available data, in the range from $\sqrt{s}$=19GeV to 1800GeV, leads to the conclusion that $B_R > B_I$, although the determination of $B_R$ is far 
less precise than $B_I$, for obvious reasons. Note that strickly speaking, this concept of hadronic slope is very misleading, since
it is known that the derivative of the amplitude with respect to $|t|$ is a slowly decreasing function of $|t|$ as shown in Ref.~\cite{BSW84}, an approach where
real and imaginary parts of the amplitude are strongly related. The relevance of the measurement of the real part of the $pp$ forward
scattering amplitude at the LHC has been also emphasized in Ref.~\cite{BKMSW}.\\ 
A model for $pp$ and $\bar pp$ elastic scattering, based on the electromagnetic and gravitational form factors related to a new set of generalized parton distributions (GPD) 
was used, after unitarization, to fit the data~\cite{OS}. Unfortunately the quality of the fit is rather poor, with a $\chi^2$/pt=6 and it predicts a high value of the total cross section at LHC, $\sigma_{tot}$=146mb. One gets an even higher prediction at $\sqrt{s}$=14TeV, $\sigma_{tot}$=230mb, in another approach, which introduces the concept of reflective elastic scattering at very high energies~\cite{ST}. This picture also predicts that the
scattering amplitude at the LHC energy goes beyond the black disk limit.

Another phenomenological investigation of $pp$ and $\bar pp$ elastic scattering was carried out by considering that the proton consists of an outer
region of $\bar qq$ condensed ground state, an inner shell of topological baryonic charge and a core where valence quarks are confined~\cite{RL}. It
leads to $\sigma_{tot}$=110mb and for the ratio of real to imaginary parts of the forward amplitude $\rho$=0.12 at LHC. The predicted differential cross section $d\sigma/dt$ has a smooth behavior beyond the bump at $|t|\simeq1\mbox{GeV}^2$, with no oscillations and a much larger value, in contrast with other models.

Concerning the specific issue of the value of the $pp$ total cross section at the LHC, a highly non-perturbative quantity which cannot be predicted by QCD, we had a general presentation of different models (double poles, triple poles, cuts, etc...) and their experimental consequences~\cite{JRC}. It was stressed that the theoretical uncertainty is large, as discussed above, and therefore an accurate measurement is badly needed since it will also
tell us a lot about the analytic structure of the $pp$ elastic amplitude.

The eikonal approach has been proven to be very useful in describing high energy elastic scattering. Clearly it relies on the knowledge of the impact parameter profile, which can be either more peripheral or central. The analysis of the $pp$ data at the ISR energy $\sqrt{s}$=53GeV led to the conclusion that a peripheral profile is preferred in this case~\cite{VK}. In another presentation~\cite{MVL}, the validity of the optical theorem commonly used to extract the total cross section has been questionned. 

A new rigorous result on the inelastic cross section was obtained recently~\cite{AM} and it reads $\sigma_{inel}(s)< \pi/4m_{\pi}^2 (\mbox{ln} s)^2$. This bound is four times smaller than the old Froissart bound derived in 1967, $\sigma_{tot}(s)< \pi/m_{\pi}^2 (\mbox{ln} s)^2$ where $\pi/m_{\pi}^2$ = 60mb. This last result can be also improved by a factor two, using some reasonable assumptions and it would be nice to prove it rigorously.

A possible description of high-energy small-angle scattering in QCD can be done by means of two vacuum exchanges with $C=\pm 1$, the Pomeron and the 
Odderon. Recent developments in this subject, based on the weak/strong duality, relating Yang-Mills theories to string theories in Anti-de Sitter (AdS)
space, were presented in some details~\cite{CIT}. If the QCD Pomeron is viewed as a two-component object, soft and hard, a dual description of the Pomeron emerges unambiguously through the AdS/CFT approach and the Odderon is related to the anti-symmetric Kalb-Ramond field. Some aspects of
analyticity, unitarity and confinement were also discussed.

\section{Soft Diffraction}
In an overview of soft diffraction~\cite{AK}, several theoretical approaches were considered for
a better understanding of the relevant mechanisms of high-energy interactions and making an instructive comparison between s- and t-channel
view points. Diffractive production in the s-channel is peripheral in the impact parameter and there is
a strong influence of unitarity effects due to multi-pomeron exchanges.
The calculation of the survival probability for hard processes is very important, in particular for high mass diffraction, as we will see later, for
example, for central Higgs production.

Following the above ideas, a model based on Gribov's Reggeon calculus was proposed and applied to soft diffraction processes at high energy~\cite{MP}. By giving a special attention to the absorptive corrections, the parameters of the model are determined following from a good description of the existing experimental data on inclusive diffraction in the energy range from ISR up to Tevatron. The model predictions for single and double diffraction at LHC energy are also given later. In another contribution~\cite{EM}, one was recalling the method to unitarize the Pomeron for elastic
and inclusive scattering, providing as well as a comparison with data, mainly for one-particle inclusive production and some LHC predictions.

Soft scattering theory was re-visited by considering some eikonal models for simplicity and to secure s-channel unitarity~\cite{UM}. After
recalling the main features of two specific models~\cite{GLMM,RMK}, the interplay between theory and data analysis led to some LHC predictions, in
particular a total cross section of the order of 90mb, in contrast with the prediction $\sigma_{tot}=103.6 \pm 1.1$mb from Ref.~\cite{BSW}. Another
important point from Ref.~\cite{BSW} to notice here, is the fact that the ratio $\sigma_{el}/\sigma_{tot}$ rises from the value 0.18
for $\sqrt{s}$=100GeV to 0.30 for $\sqrt{s}$=100TeV, whereas Refs.~\cite{GLMM,RMK} predict almost no energy dependence in this range.

Some special features of the model of soft interactions of Ref.~\cite{GLMM} mentioned above, were discussed together with the
results of the fit to determine the parameters of the model~\cite{EG}. Needless to say that it is very important to estimate the survival probability for central exclusive production of the Higgs boson, which was also compared with the results of the model of Ref.~\cite{RMK}.

This question is also related to the notion of color fluctuations in the nucleon in high energy scattering~\cite{MS}, so it is legitimate to ask: how
strong are fluctuations of the gluon field in the nucleon? A simple dynamical model can explain the ratio of the inelastic to the elastic cross section in vector meson production in $ep$ collisions at HERA and leads to a new sum rule~\cite{FSTW}. However it cannot explain the Tevatron CDF data and it reduces the expected survival probability in central exclusive production. 

\section{Hard Diffraction and Central Production}
The standard QCD mechanism for central exclusive production for heavy systems, using the formalism of collinear generalized parton 
distributions has been proposed some time ago and applied for Higgs production at the LHC~\cite{KKMR}. In this case also, it is relevant to question a possible violation of QCD factorization and some aspects of analyticity and crossing properties~\cite{OT}. At the phenomenological level, the
same mechanism was used to calculate the amplitudes for the central exclusive production of the $\chi_c$ mesons, using different unintegrated gluon 
distribution functions (UGDF)~\cite{TPS}. The extention of the UGDF to the non-forward case, can be obtained by saturation of positivity constraints. The resulting total cross sections for all charmonium states $\chi_c(0^+,1^+,2^+)$ are compared at Tevatron energy.
 
The present situation of theoretical predictions for central exclusive production of Higgs bosons and other heavy systems at
the LHC was reviewed~\cite{JRC2}. It was shown that the CDF dijet data can be used to reduce the uncertainty on the cross section prediction for
the Higgs boson. The claim is that a cross section between 0.3 and 2fb is expected for a standard Higgs of mass 120GeV. Central exclusive production of vector mesons may be used as a discovery channel for the odderon.

Some simple examples of physics beyond the Standard Model (SM), which require an extended Higgs sector, were considered~\cite{SH}. Assuming a central
exclusive production mechanism, the sensitivity of the search for the corresponding Higgs was studied, with some experimental aspects like signal
and background rates. In another presentation, along the same lines of extending the Higgs sector beyond the SM, the search for the lightest neutral Higgs boson of a model containing triplets was discussed~\cite{KH}. By means of some Monte Carlo simulations, it was found that the central exclusive production mechanism is again a very powerful tool to study this new object. 

Deep-inelastic scattering data in the very low-$x$ region is known to be dominated, in the Regge picture, by the Pomeron. By using 
a discretized version of the BFKL Pomeron, which generates discrete Regge pole solutions, an integrated positive gluon distribution was obtained~\cite{DR}. It allows a good fit of the ZEUS $F_2$ data in the kinematic range $10^{-4}<x<10^{-2}$ and $4.5< Q^2<350\mbox{GeV}^2$ and
this gluon distribution must be tested in hadronic collisions at the LHC.

In jet production at LHC, gaps between jets is an important issue which deserves serious theoretical studies, because, it is
sensitive to various QCD processes. The phenomenological impact of the Coulomb gluon contributions and super-leading logarithms on the gaps
between jets cross section, has been investigated~\cite{SM}.\\

\section{Acknowledgments}
I would like to thank the conference organizers for their warm hospitality
at CERN and for making the 13th "`Blois Workshop"' a very successful meeting. I am
also grateful to all the conference speakers for the high quality of their contributions.

\end{document}